\documentclass[conference]{IEEEtran}
\IEEEoverridecommandlockouts
\usepackage{cite}

\usepackage{graphicx}
\usepackage{tikz}
\usetikzlibrary{shapes.geometric}
\usetikzlibrary{positioning,calc,arrows.meta}

\usepackage{float}

\usepackage{geometry}
\usepackage{array}          
\usepackage{multirow}       
\usepackage{makecell}       

\usepackage{booktabs}       
\usepackage{longtable}      
\usepackage{tabularx}       
\usepackage{rotating}       
\usepackage{array}

\usepackage{graphicx}       
\usepackage{microtype}

\usepackage{pifont}         

\usepackage{enumitem}       
\usepackage{tcolorbox}      
\tcbuselibrary{skins,breakable}

\usepackage{hyperref}   

\usepackage{amsmath,amssymb}
\usepackage{algorithm}   
\usepackage{algpseudocode}  
\usepackage{float}

\algrenewcommand\algorithmicrequire{\textbf{Inputs:}}
\algtext*{EndFor}

\usepackage{esvect}
\usepackage{bm}
\newcommand{\vx}{\mathbf{x}}         

\newcommand{\Loss}{L}                
\newcommand{\Leakage}{\mathrm{Leakage}} 
\newcommand{\CVaR}{\mathrm{CVaR}}    

\def\BibTeX{{\rm B\kern-.05em{\sc i\kern-.025em b}\kern-.08em
    T\kern-.1667em\lower.7ex\hbox{E}\kern-.125emX}}

\begin{document}



\title{Towards a  Risk-Cost Model for Financial Adaptive Authentication}


\author{\IEEEauthorblockN{Supriya Khadka}
\IEEEauthorblockA{\textit{George Mason University} \\
Fairfax, Virginia, USA \\
skhadk@gmu.edu}
\and
\IEEEauthorblockN{Sanchari Das}
\IEEEauthorblockA{\textit{George Mason University} \\
Fairfax, Virginia, USA \\
sdas35@gmu.edu}
}

\maketitle


\begin{abstract}
Authentication in financial systems remains a uniquely high-stakes security challenge, where even marginal increases in false acceptance can result in catastrophic monetary loss. Existing deployments of adaptive authentication, which combine biometrics, behavioral signals, and contextual risk scoring, remain conceptually fragmented and often prioritize regulatory compliance over explicit economic and adversarial risk modeling. To address this structural imbalance, in this paper we introduce a formal \textit{Risk-Cost Model (RCM)} for adaptive authentication in financial systems. The RCM provides a principled mathematical foundation that integrates three essential components: (i) cost-sensitive risk functions that explicitly capture fraud loss, opportunity cost, and tail risk through Conditional Value-at-Risk ($\mathrm{CVaR}$); (ii) sequential decision-making mechanisms that adapt to adversarial probing and distributional drift; and (iii) quantifiable privacy and regulatory constraints embedded directly within the optimization objective. By reframing authentication as a constrained dynamic risk-cost optimization problem, the RCM moves beyond static classification and compliance-driven design toward systems that are economically grounded, tail-risk aware, and resilient under adversarial uncertainty. 
\end{abstract}


\section{Introduction}
Authentication in financial and cryptocurrency systems represents a uniquely high-stakes security control~\cite{taher2019enhanced, singh2025securing, khadka2026grant,das2020risk,das2019towards}. In such environments, transactions are inherently both liquid and irreversible, so even marginal increases in false acceptance rates can translate into catastrophic monetary loss~\cite{kutera2022cryptocurrencies,podapati2025sok}. Traditionally, authentication has relied on static mechanisms such as passwords, PINs, and cryptographic keys, which operate through deterministic checks or fixed thresholds~\cite{yusop2025advancing, nandy2019review, barkadehi2018authentication,das2020mfa,noah2025pins}. However, repeated compromise incidents demonstrate that static authentication is vulnerable to adversaries who probe systems, exfiltrate credentials, and exploit phishing or credential-stuffing attacks~\cite{oosthoek2021cyber, kothamasu2023investigation,das2020smart,majumdar2021sok,noah2025replication,das2019evaluating}.

In response to these evolving threats, financial platforms are turning to adaptive authentication~\cite{podapati2025sok,kishnani2024dual}. Rather than making a single binary decision based on static inputs, adaptive systems dynamically integrate multiple modalities, including biometrics, behavioral signals, and contextual metadata, to calibrate authentication challenges according to estimated risk~\cite{bakar2014adaptive, chitraju2025ai, bumiller2023understanding, oduri2024continuous}. The objective is to strike a principled balance between fraud prevention and user convenience. Yet, despite this promise, existing deployments remain conceptually fragmented. Many current frameworks emphasize privacy and regulatory compliance, but they often lack explicit modeling of monetary loss and rarely account for adversarial dynamics over time~\cite{nzomiwu2025cybersecurity, umakor2022threat}.

Thus we see, current approaches to financial adaptive authentication frequently emphasize privacy and regulatory compliance, yet often lack explicit economic modeling and rigorous treatment of adversarial dynamics. Consequently, authentication mechanisms are commonly implemented as refined classification systems rather than as principled decision processes that directly optimize financial risk~\cite{barkadehi2018authentication, chenchev2021authentication}. This observation builds on insights from our prior systematic analysis of the financial adaptive authentication landscape (under review). 

To address this limitation, we introduce a formal \textit{Risk-Cost Model (RCM)} for financial adaptive authentication. The RCM reconceptualizes authentication as a constrained dynamic optimization problem in which each authentication attempt is evaluated through cost-sensitive risk functions that explicitly capture fraud loss, opportunity cost, and tail risk via $\mathrm{CVaR}$. The model further incorporates sequential adaptation to mitigate adversarial probing and distributional drift, while embedding privacy and regulatory constraints directly within the optimization objective. By integrating economic risk modeling, adversary-aware sequential decision-making, and regulatory considerations into a unified mathematical framework, the RCM provides a principled foundation for designing authentication systems that are not only compliant but economically resilient and strategically strong.

\section{Related Work}
Adaptive authentication has a long history in usable security and identity systems, typically framed as \emph{risk-based authentication} that fuses device, network, behavioral, and contextual signals to trigger step-up checks when estimated risk is high~\cite{oosthoek2021cyber, bakar2014adaptive}. Prior surveys systematize the design space of adaptive authentication by cataloging signals, fusion strategies, and policy logic, and by highlighting the usability--security trade-offs that arise in real deployments~\cite{arias2019survey,preuveneers2021authguide,ryu2023design, das2019evaluating}. While these surveys provide broad coverage of mechanisms, they largely treat decision-making as thresholding or classification and do not explicitly formalize authentication as an economic optimization problem under adversarial dynamics in financial settings.

\subsection{Financial and cryptocurrency authentication.}
Financial and cryptocurrency platforms amplify the consequences of authentication error because transactions are often irreversible and assets are liquid, making false accepts directly translate to monetary loss~\cite{taher2019enhanced,kutera2022cryptocurrencies, khadka2026poster}. Empirical and forensic work on compromises in online services and crypto ecosystems underscores that credential theft, replay, and account takeover remain persistent drivers of loss~\cite{oosthoek2021cyber,kothamasu2023investigation}. As a result, recent work has explored deploying adaptive authentication in banking and payment systems using behavioral signals and transaction context~\cite{bakar2014adaptive,abuhamad2020sensor,chitraju2025ai, podapati2025sok}, while also highlighting the security, privacy, and accessibility gaps present in widely deployed e-payment and MFA applications~\cite{kishnani2023assessing, kishnani2024towards, jensen2021multi}. However, many proposals emphasize detection accuracy and operational heuristics, with limited attention to explicit \emph{risk-cost} objectives that represent fraud loss, user opportunity cost, and challenge friction within a single decision criterion.

\definecolor{RCMblueRGB}{RGB}{0,114,178}
\definecolor{RCMgreenRGB}{RGB}{0,158,115}
\definecolor{RCMredRGB}{RGB}{213,94,0}
\definecolor{RCMamberRGB}{RGB}{230,159,0}
\definecolor{RCMinkRGB}{RGB}{40,40,40}

\newif\ifRCMbw

\ifRCMbw
  \colorlet{RCMblue}{black!15}
  \colorlet{RCMgreen}{black!25}
  \colorlet{RCMred}{black!30}
  \colorlet{RCMamber}{black!40}
  \colorlet{RCMink}{black}
\else
  \colorlet{RCMblue}{RCMblueRGB}
  \colorlet{RCMgreen}{RCMgreenRGB}
  \colorlet{RCMred}{RCMredRGB}
  \colorlet{RCMamber}{RCMamberRGB}
  \colorlet{RCMink}{RCMinkRGB}
\fi

\begin{figure*}[t]
\centering
\resizebox{\textwidth}{!}{%
\begin{tikzpicture}[
  >=Latex, line cap=round,
  node distance=9mm and 12mm,
  every node/.style={font=\scriptsize, align=center, transform shape, text=RCMink},
  Route/.style={-{Latex[length=2.6mm,width=2.2mm]}, line width=1.5pt},
  HaloLabel/.style={midway, fill=white, rounded corners=2pt, inner sep=1pt, text=RCMink},
  chip/.style={fill=white, rounded corners=2pt, inner sep=1pt},
  chiptiny/.style={fill=white, rounded corners=1pt, inner sep=0.6pt},
  theory/.style={draw=RCMink, fill=white, rounded corners=2pt, inner sep=2pt},
  box/.style={draw=RCMblue!70!black, rounded corners=3pt, thick, fill=RCMblue!7, inner sep=5pt, text width=31mm},
  decision/.style={draw=RCMgreen!60!black, rounded corners=3pt, line width=1pt,
                   double=RCMgreen!30, double distance=1.6pt, fill=RCMgreen!10,
                   inner sep=6pt, text width=36mm},
  risk/.style={draw=RCMred!65!black, ellipse, line width=0.9pt, fill=RCMred!10,
               inner sep=2pt, minimum width=22mm, text width=26mm},
  concept/.style={draw=RCMamber!60!black, rounded corners=3pt, line width=0.9pt,
                  fill=RCMamber!9, inner sep=6pt, text width=44mm},
  conceptWide/.style={concept, text width=48mm},
  dot/.style={circle, fill=RCMink, inner sep=0.8pt},
  step/.style={circle, draw=RCMink, fill=white, inner sep=1pt, font=\footnotesize\bfseries}
]

\node[box] (features) {\textbf{Input Features}\\[2pt]
$\mathbf{x}\!\in\!\mathcal{X}$\\
Contextual + behavioral signals};

\node[box, below=of features] (prob) {\textbf{Calibrated Probability}\\[2pt]
$p_i(\mathbf{x})=\Pr(Y=i\mid \mathbf{x})$\\
Maps features $\to$ impostor likelihood};

\node[above=1.2mm of features, font=\bfseries] {Context \& Evidence};

\node[decision, right=of features, yshift=-5mm] (policy) {\textbf{Decision Policy}\\[1.5pt]
$a\in\{\textsc{accept},\textsc{challenge},\textsc{reject}\}$\\
\emph{Minimize expected cost}};

\node[below=1.5mm of policy, font=\bfseries] {Decision Space};

\node[theory, anchor=south] (bayes) at ($(policy.south)+(0mm,20mm)$) {%
\(\displaystyle \text{accept if } \frac{p_i(\mathbf{x})}{1-p_i(\mathbf{x})}
< \frac{c_{\mathrm{FR}}}{c_{\mathrm{FA}}}\)};
\draw[dashed, draw=RCMink] (bayes.south) -- ++(0,-3mm);

\node[risk, right=19mm of policy, yshift=14mm] (fa) {\textbf{False Accept}\\Loss $c_{\mathrm{FA}}$\\Fraud exposure};
\node[risk, right=19mm of policy] (fr) {\textbf{False Reject}\\Loss $c_{\mathrm{FR}}$\\User churn};
\node[risk, right=19mm of policy, yshift=-14mm] (ch) {\textbf{Challenge}\\Cost $c_{\mathrm{CH}}(\mathbf{x})$\\User friction};

\node[above=1.2mm of fa, font=\bfseries] {Risk Components};

\node[chiptiny, anchor=west] at ($(ch.west)+(-6mm,-3mm)$) {$\rho(\mathbf{x})$};

\node[concept, right=22mm of fr] (riskcomb) {\textbf{Risk–Cost Functional}\\[3pt]
$\mathcal{R}(d)=
c_{\mathrm{FA}}\mathrm{FAR}+
c_{\mathrm{FR}}\mathrm{FRR}+
c_{\mathrm{CH}}\mathrm{CHR}+
\lambda\,\mathrm{Leakage}(d)$\\[2pt]
\emph{Unifies fraud, friction, privacy}};

\node[concept, below=10mm of riskcomb] (cvar) {\textbf{Tail Risk (CVaR)}\\[2pt]
$\displaystyle \mathrm{CVaR}_\alpha(L)=
\inf_t\Big\{ t+\tfrac{1}{1-\alpha}\,\mathbb{E}\big[(L-t)_+\big]\Big\}$\\
\emph{Typical choice: }\(\alpha\in[0.95,0.999]\)};

\node[conceptWide, below=10mm of cvar] (robust) {\textbf{Adversary–Aware Design}\\[2pt]
$\displaystyle \min_d\ \sup_{Q\in\mathcal{U}(P;\delta)}
\mathbb{E}_Q\!\big[L(d(\mathbf{x}),Y)\big]+\lambda\,\mathrm{Leakage}(d)$\\
\(\mathcal{U}\): e.g., Wasserstein/TV/$\chi^2$ ball of radius $\delta$};

\node[concept, left=12mm of robust] (seq) {\textbf{Sequential Policy Optimization}\\[2pt]
$\displaystyle \min_d\ \mathbb{E}\!\Big[\sum_t L_t\Big]
+\beta\,\mathrm{CVaR}_\alpha\!\Big(\sum_t L_t\Big)
+\lambda\sum_t\epsilon_t$\\
\emph{Bandits / POMDPs for drift \& exploration}};

\node[above=1.2mm of riskcomb, font=\bfseries] {Aggregated Objectives};

\draw[Route, draw=RCMblue!80!black]
  ([xshift=1mm,yshift=1mm]features.east)
    .. controls ($(features.east)!0.5!(policy.west)+(0,10mm)$)
    .. ([xshift=-2mm,yshift=1mm]policy.west);
\node[chip, text=RCMink]
  at ($(features.east)!0.6!(policy.west)+(0,29pt)$) {signal extraction};

\draw[Route, draw=RCMblue!80!black]
  (prob.east) -- ++(6mm,0) |- ([xshift=-1.6mm,yshift=-2mm]policy.west)
  node[pos=0.22, below, chip] {calibration};

\draw[Route, draw=RCMgreen!65!black] (policy.east) -- (fa.west);
\draw[Route, draw=RCMgreen!65!black] (policy.east) -- (fr.west);
\draw[Route, draw=RCMgreen!65!black] (policy.east) -- (ch.west);

\path (riskcomb.west) ++(0, 6mm) coordinate (rcmInTop)
                      ++(0,-6mm) coordinate (rcmInMid)
                      ++(0,-6mm) coordinate (rcmInBot);

\draw[Route, draw=RCMred!80!black, shorten <=1.2pt]
  (fa.east) -- (rcmInTop) node[midway, above=3pt, chiptiny] {FAR};
\draw[Route, draw=RCMred!80!black, shorten <=1.2pt]
  (fr.east) -- (rcmInMid) node[midway, above=1.5pt, chiptiny] {FRR};
\draw[Route, draw=RCMred!80!black, shorten <=1.2pt]
  (ch.east) -- (rcmInBot) node[midway, below=3pt, chiptiny] {CHR};

\draw[Route, draw=RCMamber!80!black] (riskcomb.south) -- (cvar.north);
\draw[Route, draw=RCMamber!80!black] (cvar.south) -- (robust.north);
\draw[Route, draw=RCMamber!80!black] (robust.west) -- (seq.east);

\node[draw=RCMink, fill=white, rounded corners,
      font=\scriptsize, align=left, inner sep=3pt,
      below=10mm of ch.south, xshift=3mm] (voi) {Issue step-up iff\\
$\mathrm{VoI}(\mathbf{x})>c_{\mathrm{CH}}+\lambda\,\Delta\mathrm{Leakage}$};
\draw[Route, draw=RCMink]
  ([xshift=1mm,yshift=-1mm]ch.south)
    -- ([xshift=-2mm,yshift=1mm]voi.north);

\end{tikzpicture}%
}
\caption{RCM metro-map. Contextual evidence is calibrated to impostor posteriors \(p_i(\mathbf{x})\) (blue) and routed to a Bayes decision hub (green), which accepts when \(\tfrac{p_i(\mathbf{x})}{1-p_i(\mathbf{x})}<\tfrac{c_{\mathrm{FR}}}{c_{\mathrm{FA}}}\) and triggers a step-up when \(\mathrm{VoI}(\mathbf{x})>c_{\mathrm{CH}}(\mathbf{x})+\lambda\,\Delta\mathrm{Leakage}\). The three loss channels (red) FAR, FRR, CHR feed the risk functional \(\mathcal{R}(d)=c_{\mathrm{FA}}\mathrm{FAR}+c_{\mathrm{FR}}\mathrm{FRR}+c_{\mathrm{CH}}\mathrm{CHR}+\lambda\,\mathrm{Leakage}(d)\). This objective then flows to tail-risk control via \(\mathrm{CVaR}_\alpha\) (typical \(\alpha\!\in\![0.95,0.999]\)), to distributional robustness over \(\mathcal{U}(P;\delta)\), and finally to sequential policy optimization (bandits/POMDPs) that adapts thresholds under drift while respecting privacy budgets.}
\label{fig:rcm}
\end{figure*}
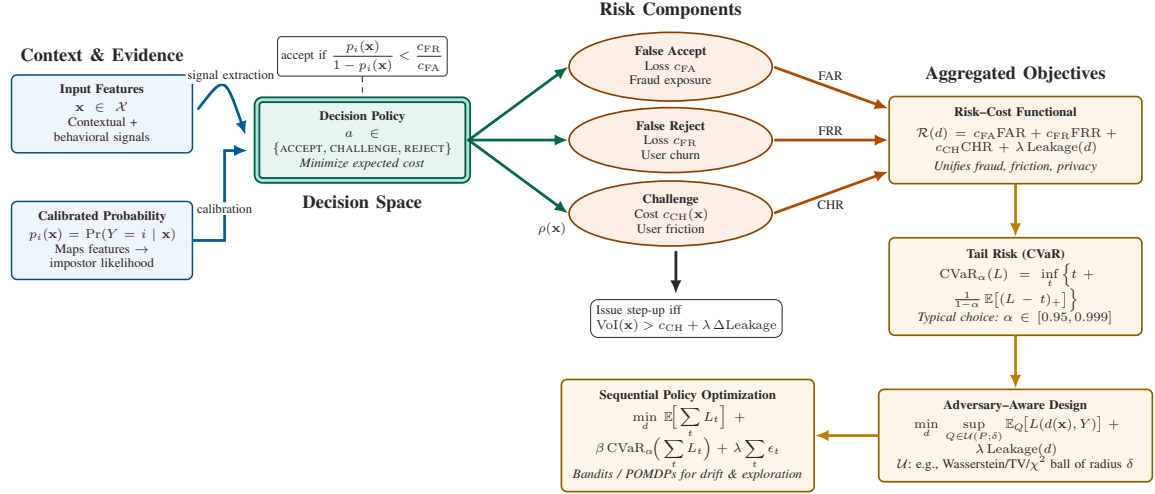

\subsection{Cost-sensitive decision-making for authentication.}
A consistent limitation across adaptive authentication studies is the absence of formal cost modeling. Many systems report FAR/FRR/EER and tune thresholds to satisfy security targets, but do not map these error rates to domain-specific costs that reflect fraud loss and business impact~\cite{oosthoek2021cyber,abuhamad2020sensor}. Cost-sensitive learning and statistical decision theory provide a natural foundation for such mapping, but they are not commonly applied end-to-end in financial authentication pipelines. In contrast, the Risk-Cost Model (RCM) developed in this paper treats authentication as a decision problem in which actions (accept, challenge, reject) are selected by minimizing expected loss under explicit cost parameters, enabling principled operating-point selection and transparent trade-offs.

\subsection{Sequential adaptation and adversarial dynamics.}
Authentication is inherently repeated: users (and attackers) interact with systems over time, during which behavior drifts and adversaries probe policies. Prior work discusses adaptive thresholds and risk scoring under changing contexts~\cite{oosthoek2021cyber,bakar2014adaptive}, and emerging systems incorporate learning-based components for continuous verification~\cite{chitraju2025ai}. Nevertheless, fully sequential formulations (e.g., online decision-making with feedback) remain relatively uncommon, and adversarial probing is often treated implicitly rather than modeled as a strategic interaction~\cite{oosthoek2021cyber}. The RCM explicitly accommodates sequential decisions by defining cumulative loss objectives and allowing policies to update under drift and probing, aligning adaptive authentication with online optimization perspectives while remaining grounded in interpretable cost terms.

Existing research provides (i) broad taxonomies and surveys of adaptive authentication mechanisms~\cite{arias2019survey,ryu2023design,oosthoek2021cyber}, and (ii) domain-specific evidence that financial systems demand stronger, more adaptive defenses~\cite{taher2019enhanced,kutera2022cryptocurrencies}. This paper complements these lines by introducing a unified Risk-Cost Model that directly connects calibrated risk estimation to cost-sensitive action selection, extends naturally to sequential decision-making under drift and probing, and embeds privacy/regulatory considerations within the same optimization framework.

\section{The Risk-Cost Model (RCM)}
Risk-cost modeling provides the quantitative foundation for adaptive authentication in financial systems by explicitly aligning authentication decisions with expected economic loss, user friction, and privacy exposure. The RCM metro map is shown in Figure~\ref{fig:rcm}.

The optimization flow proceeds sequentially: first, contextual evidence is calibrated into impostor probabilities. Second, the Bayes decision policy evaluates the expected risk across available actions to minimize the immediate one-step Risk-Cost Functional $\mathcal{R}(d)$. Finally, these immediate decisions feed into higher-order, long-term optimizations, specifically tail-risk control (CVaR) and sequential policy updating under adversarial drift.

\subsection{Problem Setup and Decision Space}
Each authentication attempt is formulated as a binary hypothesis test with an enriched action space defined as $a \in \mathcal{A} = \{\text{ACCEPT, CHALLENGE, REJECT}\}$. A CHALLENGE represents a step-up verification mechanism, such as requesting a biometric scan, an SMS one-time password (OTP), or a push notification. This action gathers additional signal evidence but incurs a measurable cost in user friction.

A given feature vector $x \in \mathcal{X}$ is mapped to a calibrated impostor probability $p_i(x) = Pr(Y=i|x)$, where $Y \in \{u,i\}$ denotes whether the user is legitimate or an impostor. The system's policy, denoted as $d: \mathcal{X} \rightarrow \Delta(\mathcal{A})$, defines a distribution over actions given the observed features.

The core of this model is the cost-sensitive one-step risk functional, which embeds financial, usability, and regulatory considerations directly into the decision process:

\begin{equation*}
\label{eq:one-step-risk}
\begin{split}
\mathcal{R}(d) \;=\;&\;
c_{\mathrm{FA}}\cdot \mathrm{FAR}(d)
\;+\; c_{\mathrm{FR}}\cdot \mathrm{FRR}(d) \\[3pt]
&\;+\;
c_{\mathrm{CH}}\cdot \mathrm{CHR}(d)
\;+\; \lambda \cdot \mathrm{Leakage}(d).
\end{split}
\end{equation*}

In this equation, $c_{FA}$ represents the expected fraud loss from false accepts, $c_{FR}$ captures the opportunity cost from false rejects, $c_{CH}$ measures user friction caused by challenges~\cite{das2018johnny, das2019mfa}, and $Leakage(d)$ quantifies the privacy or regulatory exposure.

For practical implementation, $Leakage(d)$ can be quantified as a discrete penalty, such as the number of personally identifiable information attributes exposed during a step-up challenge, or as a formal information-theoretic metric like differential privacy loss ($\epsilon$). The parameter $\lambda$ serves as the shadow price of privacy, converting this exposure into a common cost metric.

\subsection{Bayes-Optimal Action and Expected Risks}
Assuming $p_i(x)$ is well-calibrated and a challenge succeeds with probability $\rho(x)$ at an incremental cost of $c_{CH}(x)$, the expected risks for each action are evaluated as follows:

{\small
\begin{align*}
\mathsf{Risk}(\textsc{accept}\mid \mathbf{x}) 
&= p_i(\mathbf{x})\,c_{\mathrm{FA}},\\
\mathsf{Risk}(\textsc{reject}\mid \mathbf{x}) 
&= (1-p_i(\mathbf{x}))\,c_{\mathrm{FR}},\\
\mathsf{Risk}(\textsc{challenge}\mid \mathbf{x}) 
&= c_{\mathrm{CH}}(\mathbf{x}) 
+ p_i(\mathbf{x})\,(1-\rho(\mathbf{x}))\,c_{\mathrm{FA}}\\
&\quad 
+ (1-p_i(\mathbf{x}))\,(1-\rho(\mathbf{x}))\,c_{\mathrm{FR}}.
\end{align*}
}

If challenges are excluded from the action space, the optimal decision reduces to a posterior-odds threshold where the system accepts the user if:

\[
\frac{p_i(\mathbf{x})}{1-p_i(\mathbf{x})} < \frac{c_{\mathrm{FR}}}{c_{\mathrm{FA}}}.
\]

\subsection{Calibration and Stability}
Reliable cost modeling requires that scores are both calibrated and stable over time. Calibration can be achieved by mapping raw classifier scores to well-formed probabilities using isotonic regression or Platt scaling, with reliability diagrams serving as a diagnostic for probability alignment.

Once calibrated, cost curves provide a way to plot expected loss against decision thresholds under domain-specific parameters ($c_{\mathrm{FA}}, c_{\mathrm{FR}}, c_{\mathrm{CH}}$), enabling the identification of operating points that minimize loss. Long-term deployment further demands robustness to drift; stability indices and expected loss must be monitored over rolling windows, and thresholds should be adjusted dynamically to maintain compliance with risk and cost budgets

\subsection{Tail Risk and Multi-Step Aggregation}
Average-case evaluation is insufficient for financial authentication because a single breach can yield catastrophic consequences. To account for rare high-loss events, the model incorporates Conditional Value-at-Risk ($CVaR$) at a confidence level $\alpha$. Here, $L$ represents a random variable denoting the continuous monetary loss distribution associated with the authentication policy:

\[
\mathrm{CVaR}_\alpha(L)=\inf_{t\in\mathbb{R}}\Big\{t+\tfrac{1}{1-\alpha}\,\mathbb{E}\big[(L-t)_+\big]\Big\}.
\]

This formulation follows the foundational optimization framework for Conditional Value-at-Risk established by Rockafellar and Uryasev~\cite{rockafellar2000optimization}.

For multi-step authentication, policies are evaluated by combining expected loss, tail risk, and privacy exposure.

\begin{equation*}
\label{eq:cvar-objective}
\min_{d\in\mathcal{D}} \; \mathbb{E}\!\Big[\sum_{t=1}^T L_t\Big]
\;+\; \beta \cdot \mathrm{CVaR}_\alpha\!\Big(\sum_{t=1}^T L_t\Big)
\;+\; \lambda \sum_{t=1}^T \epsilon_t,
\end{equation*}

This formulation reflects the fact that systemic breaches, although infrequent, dominate financial losses.

\subsection{Adversary-Aware Design}
Authentication systems must anticipate adaptive adversaries because attackers can probe the system, exploit leaked credentials, or adapt behavior over time, making nominal evaluations insufficient~\cite{krishan2024adversarial}. The RCM captures this by embedding robustness directly into the optimization, ensuring that the policy performs well even under worst-case plausible distributions within an uncertainty set.

\[
\begin{aligned}
\min_{d\in\mathcal{D}} \;
&\sup_{Q\in\mathcal{U}(P;\delta)}
\mathbb{E}_{(\mathbf{x},Y)\sim Q}\!\big[L(d(\mathbf{x}),Y)\big]\\[2pt]
&\quad + \lambda\,\mathrm{Leakage}(d).
\end{aligned}
\]

Here, $\mathcal{U}(P;\delta)$ is an uncertainty set (for example, an f-divergence or Wasserstein ball) around the nominal distribution $P$. This distributionally robust formulation guarantees that authentication performance remains bounded despite adversarial probing, distributional shifts, or leaks, while still accounting for privacy considerations.

\subsection{Value of Information and Step-Up Challenges}
Triggering a step-up challenge can be framed as a value-of-information problem. Let $\Delta(x)$ denote the net gain from observing an additional signal $Z$. The expected value is

\[
\begin{aligned}
\mathrm{VoI}(\mathbf{x}) 
&= \min_a \mathsf{Risk}(a\mid \mathbf{x}) - \mathbb{E}\!\left[\min_a \mathsf{Risk}(a\mid \mathbf{x},Z)\right]\\[2pt]
&\quad
- \big(c_{\mathrm{CH}}(\mathbf{x})+\lambda\,\Delta\mathrm{Leakage}\big).
\end{aligned}
\]

A step-up challenge is issued only when its expected informational benefit exceeds the associated user friction and privacy cost.

\subsection{Privacy and Compliance Constraints}
Privacy and regulatory requirements can be incorporated either as hard constraints that restrict feasible policies or as soft penalties through the leakage term in the primary risk functional~\cite{das2019privacy, adhikari2025policypulse, markert2023transcontinental, adhikari2023evolution,goswami2026sok}. A Lagrangian approach allows exploring a range of operating points that balance fraud prevention with privacy and compliance obligations.

\subsection{Sequential Policy Optimization}

Because authentication unfolds over time, risk-cost modeling naturally extends into a sequential decision framework, allowing policies to adapt with new observations. At each time step $t$, the system observes features $x_t$ and calculates expected one-step risk $R(a_t|x_t)$ for each action $a_t \in \{ACCEPT, CHALLENGE, REJECT\}$. If an action triggers a step-up challenge, it incurs an incremental privacy leakage $\Delta\epsilon_t$. The system then selects the action that minimizes the regularized sequential objective, tracking both cumulative realized loss $L_t(a_t)$ and updated privacy budget $\epsilon_t$.

This continuous process is formalized in Algorithm~\ref{alg:sequential_rcm}, which employs risk-sensitive sequential updates to adapt decision thresholds under distributional drift while rigorously maintaining privacy budgets. By minimizing the $CVaR$-regularized objective, the algorithm actively manages adversarial uncertainty and bounds potential tail-risk exposures

\begin{algorithm}[htbp]
\caption{Risk and Privacy-Aware Adaptive Authentication}
\label{alg:sequential_rcm}
\begin{algorithmic}[1]
\State \textbf{Inputs:} $c_{\mathrm{FA}}, c_{\mathrm{FR}}, c_{\mathrm{CH}}$, privacy weight $\lambda$, CVaR level $\alpha$, robust radius $\delta$
\For{$t = 1$ to $T$}
  \State Observe features $\mathbf{x}_t$
  \State Compute impostor probability $p_i(\mathbf{x}_t)$ and leakage increment $\Delta\epsilon_t$
  \State Compute action risks $R_{\mathrm{acc}}, R_{\mathrm{rej}}, R_{\mathrm{chal}}$
  \State Choose $a_t$ minimizing $\widehat{R}(a\mid\mathbf{x}_t) + \beta\,\widehat{\mathrm{CVaR}}_{\alpha}(a)$
  \State Execute $a_t$; log realized loss $L_t$; update $\epsilon_t \gets \epsilon_{t-1}+\Delta\epsilon_t$
\EndFor
\end{algorithmic}
\end{algorithm}

\section{Discussion and Implications}
Our findings reveal a persistent structural imbalance in financial adaptive authentication research. Privacy and regulatory compliance dominate system design, while explicit modeling of fraud-loss costs and sequential adaptation under adversarial dynamics remains comparatively underdeveloped. This imbalance is not surprising given the strong regulatory pressures imposed by frameworks such as GDPR and PSD2~\cite{gdpr2016, psd22015}. However, the resulting research landscape suggests that privacy has become the primary organizing principle, often at the expense of economic and adversarial considerations.

Privacy-preserving architectures are frequently treated as implicit indicators of security. Yet privacy guarantees do not automatically translate into robustness against adaptive attackers. An authentication system can minimize data exposure and satisfy compliance requirements while still operating at a decision threshold that is economically suboptimal or vulnerable to probing. In high-stakes financial environments, where even small increases in false accept probability can produce disproportionate monetary losses, security must be evaluated not only in terms of data protection but also in terms of explicit fraud-loss exposure and operating-point economics~\cite{mabitsela2025defending}.

The limited adoption of cost-sensitive modeling further amplifies this fragility. Although fraud loss, opportunity cost, and user friction are intrinsic to financial authentication~\cite{hoppner2022instance}, relatively few works formalize these quantities within a unified risk functional. Without explicit cost parameters, threshold selection becomes heuristic and detached from monetary consequences~\cite{wai_thar_2025_threshold}. This can lead to misaligned deployments in which institutions over-correct for compliance risk while under-optimizing for financial loss, or vice versa. Embedding authentication decisions within a transparent risk-cost formulation makes such trade-offs explicit and measurable~\cite{das2021short}.

A second implication concerns temporality. Financial authentication is inherently sequential: users interact repeatedly with systems, and adversaries adapt over time. Yet much of the literature continues to evaluate authentication policies as static classifiers~\cite{mohsin2019blockchain, velasquez2018authentication}. This abstraction neglects adversarial probing, distributional drift, and cumulative tail risk. When sequential structure is ignored, systems may appear secure under one-shot evaluation while remaining vulnerable to gradual threshold inference or repeated low-value fraud that aggregates into substantial loss. Framing authentication within a dynamic optimization perspective highlights the importance of monitoring cumulative exposure and adapting policies under uncertainty.

The RCM presented in this work serves as a unifying analytical lens for interpreting these gaps. By expressing authentication as a constrained optimization problem that integrates fraud loss, opportunity cost, challenge friction, tail risk, and privacy leakage within a single objective, the RCM clarifies how design choices shift economic and regulatory trade-offs. Rather than proposing a new threat model, the RCM provides a structured foundation for evaluating existing systems and guiding future implementations toward economically coherent and adversary-aware operation.

For practitioners, the implication is straightforward: compliance alone is insufficient as a design objective. Financial authentication systems should be evaluated against explicit cost functions and monitored over time for cumulative and tail-risk exposure. For researchers, the implication is methodological: future work should integrate cost-sensitive objectives and sequential reasoning directly into evaluation protocols, rather than treating them as secondary concerns. Aligning privacy, economic loss, and adaptive decision-making within a unified risk-cost perspective is essential for building authentication mechanisms that remain resilient in adversarial financial environments.

\section{Limitations and Future Work}
This work focuses on formalizing the Risk-Cost Model (RCM), which assumes accurate estimation of cost parameters ($c_{\mathrm{FA}}, c_{\mathrm{FR}}, c_{\mathrm{CH}}$), calibrated probabilities, and measurable privacy leakage; however, these quantities may be difficult to estimate precisely in practice. Institutions can operationalize the model using proxy metrics: challenge friction cost ($c_{\mathrm{CH}}$) and user drop-off rate ($CHR(d)$) can be estimated through A/B testing of step-up authentication flows and mapped to Customer Lifetime Value (CLV) or average cart-abandonment costs; challenge success probability ($\rho(x)$) can be derived from historical authentication logs; and $Leakage(d)$ can be bounded using discrete data-exposure counts tied to institutional compliance risk matrices. We will further develop principled methods for estimating domain-specific cost parameters, implement RCM-driven policies on real-world financial workloads, and design scalable algorithms for CVaR-regularized, privacy-constrained online optimization with formal robustness guarantees.

\section{Conclusion}
Financial authentication failures translate directly into monetary loss, making economically grounded decision-making essential. In this paper, we introduced a formal \textit{Risk--Cost Model} (RCM) that treats adaptive authentication as a constrained dynamic optimization problem rather than a static classification task. We defined an explicit action space $\mathcal{A}=\{\textsc{accept},\textsc{challenge},\textsc{reject}\}$ and modeled decisions by minimizing a unified objective that internalizes fraud loss ($c_{\mathrm{FA}}$), false-reject opportunity cost ($c_{\mathrm{FR}}$), challenge friction ($c_{\mathrm{CH}}$), and privacy exposure through $\lambda \cdot \Leakage(d)$. We incorporated tail-risk control using $\CVaR_{\alpha}$ and extended the formulation to sequential settings to address drift and adversarial probing. By deriving cost-sensitive operating rules from calibrated posteriors $p_i(\vx)$ and making economic trade-offs explicit, we provide a technical foundation for designing and auditing financially robust, privacy-constrained, and adversary-aware authentication systems.

\section{Acknowledgement}
We thank Vyoma Harshitha Podapati and Harkirat Singh for their initial contributions to this work. We acknowledge the Data Agency and Security (DAS) Lab at George Mason University (GMU), where this study was conducted, and Google for partially supporting this work. The opinions expressed are solely those of the authors.


\bibliographystyle{IEEEtran}
\bibliography{rcm}

\end{document}